\documentstyle[preprint,aps]{revtex}
\begin{document}
\tightenlines
\draft
\preprint{
\parbox{4cm}{
\baselineskip=12pt
TMUP-HEL-0002\\
TIT/HEP-443\\ 
KEK-TH-682\\ 
March, 2000\\
\hspace*{1cm}
}}
\title{Models of Dynamical Supersymmetry Breaking \\
with Gauged $U(1)_R$ Symmetry}
\author{ Noriaki Kitazawa $^a$ 
  \thanks{e-mail: kitazawa@phys.metro-u.ac.jp}, 
         Nobuhito Maru $^b$ 
  \thanks{e-mail: maru@th.phys.titech.ac.jp, JSPS Research Fellow}
  \thanks{Address after April 1; Department of Physics, 
          University of Tokyo, Hongo, Bunkyo-ku, Tokyo 113-0033, Japan.}
     and Nobuchika Okada $^c$ 
  \thanks{e-mail: okadan@camry.kek.jp, JSPS Research Fellow}}  

\address{$^a$Department of Physics, Tokyo Metropolitan University,\\
         Hachioji, Tokyo 192-0397, Japan}
\address{$^b$Department of Physics, Tokyo Institute of Technology,\\
         Oh-Okayama, Meguro, Tokyo 152-8551, Japan}
\address{$^c$Theory Group, KEK, Tsukuba, Ibaraki 305-0801, Japan}
%
\maketitle
%
%
\vskip 2.5cm
\begin{center}
{\large Abstract}
\vskip 0.5cm
\begin{minipage}[t]{14cm}
\baselineskip=19pt
\hskip4mm
We present simple models of
 dynamical supersymmetry breaking with gauged $U(1)_R$ symmetry. 
The minimal supersymmetric standard model and
 supersymmetric SU(5) GUT are considered 
 as the visible sector. 
The anomaly cancellation conditions for $U(1)_R$ are investigated
 in detail and simple solutions of the R-charge assignments are found. 
We show that this scenario of dynamical supersymmetry breaking 
 is phenomenologically viable
 with the gravitino mass of order 1 TeV or 10 TeV. 
%
\end{minipage}
\end{center}
\newpage
\section{Introduction}
Supersymmety is motivated 
 to solve the gauge hierarchy problem. 
However, since the superparticles have not been observed yet, 
 supersymmetry should be broken at low energies. 
Spontaneous supersymmetry breaking at the tree-level \cite{OR-F-I}
 does not explain why the discrepancy 
 between the supersymmetry breaking scale and 
 the Planck scale is so large. 
On the other hand, 
 in the models of dynamical supersymmetry breaking 
 the supersymmetry breaking scale is related to
 the Planck scale via the dimensional transmutation \cite{Witten}. 
In the light of this fact, 
 many people have constructed the models of the dynamical 
 supersymmetry breaking \cite{dsb} and discussed
 the phenomenology so far \cite{review,GR}. 
As for the mediation mechanism of supersymmetry breaking 
 to the visible sector, there are mainly two ways. 
One is the gravity mediation \cite{review}
 (including the anomaly mediation \cite{anomaly}), 
 the other is the gauge mediation \cite{GR}
 (including the anomalous U(1) mediation \cite{anomalousU1}). 
Many authors have extensively studied both scenarios and
 proposed interesting models.

In our previous letter \cite{kitazawa},
 we proposed a simple mechanism of the dynamical
 supersymmetry breaking with gauged $U(1)_R$ symmetry.
Although the attempts to make use of the gauged $U(1)_R$
 symmetry in supergravity theories can be found
 in Refs. \cite{chamseddine},
 the mechanism of supersymmetry breaking was not the main
 topics in these works. 
Supersymmetry can be dynamically broken
 by the interplay between the Fayet-Iliopoulos term of
 $U(1)_R$ and the dynamically generated superpotential
 due to the non-perturbative dynamics of the gauge theory
 with vanishing cosmological constant.
However, we did not discuss in detail
 the anomaly cancellation for $U(1)_R$ taking into account
 the full particle contents, 
 since we focused on the dynamics of supersymmetry breaking
 in the hidden sector. 
The main purpose of this paper is to pursuit this point. 
We discuss two models as the visible sector:
 one is the minimal supersymmetric standard model (MSSM),
 the other is the supersymmetric SU(5) grand unified theory (GUT). 
In the model of MSSM, we use the Green-Schwarz mechanism
 for anomaly cancellation. 
We find quite simple solutions of R-charge assignments 
 in both cases. 
The spectrum of the supersymmetry breaking mass
 is also discussed. 
The scalar fields receive the soft supersymmetry breaking masses,
 which is the same order of the gravitino mass
 from the tree-level interactions of supergravity. 
Moreover, the scalar fields with non-zero R-charges obtain
 additional soft supersymmetry breaking masses
 through $U(1)_R$ D-term,
 which is also the same order of the gravitino mass. 
The masses of gauginos in the MSSM depend on
 the form of gauge kinetic functions. 
If the gauge kinetic function includes the higher dimensional
 term, the gaugino masses are generated in the same order
 of the gravitino mass or less. 
If the gauge kinetic function is trivial,
 the gaugino masses are generated through the anomaly mediation
 in a few orders smaller than the gravitino mass. 
We see that our scenario is phenomenologically viable
 with the gravitino mass of order 1 TeV or 10 TeV. 
The comments on the generation of the Higgs potential
 from higher dimensional interactions are given.

This paper is organized as follows. 
In section 2, 
 we review our mechanism of dynamical supersymmetry breaking 
 with gauged $U(1)_R$ symmetry. 
Then this mechanism is applied to MSSM of
 the visible sector in section 3 and 
 applied to the supersymmetric SU(5) GUT of the visible sector
 in section 4.
The last section is devoted to the summary and discussion.

\section{Dynamical Supersymmetry Breaking 
with Gauged $U(1)_R$ Symmetry}
In this section, 
we review our previous letter, 
in which we proposed a simple mechanism 
of dynamical supersymmetry breaking 
with gauged $U(1)_R$ symmetry 
in the context of the minimal supergravity. 
The model is based on the gauge group $SU(2)_H \times U(1)_R$ 
with the following matter contents.
\footnote{In our previous letter \cite{kitazawa}, 
we did not discuss 
the cancellation of the gauge anomaly of $[U(1)_R]^3$ 
and the mixed gravitational anomaly of $U(1)_R$. 
We simply assumed there that these anomalies are cancelled out, 
 if all particle contents are considered
 with an appropriate $U(1)_R$ charge assignment.
In this paper, we will discuss this issue in detail.} 
\begin{center}
\begin{tabular}{ccc}
 \hspace{1cm}& $~SU(2)_H$~ & $~U(1)_R~$  \\
$ Q_1 $   &  \bf{2}     & $ -1 $   \\
$ Q_2 $   &  \bf{2}     & $ -1 $   \\
$ S $     &  \bf{1}     & $ +4 $  
\end{tabular}
\end{center}
The general renormalizable superpotential at the tree-level is   
\begin{eqnarray}
 W = \lambda S \left[ Q_1 Q_2 \right] \; ,  
\end{eqnarray} 
 where the square brackets denote the contraction of $SU(2)$ indices 
 by the $\epsilon$-tensor, 
 $\lambda$ is a dimensionless coupling constant. 
We assume that $\lambda$ is real and positive. 

It is known that the superpotential is generated dynamically  
 by non-perturbative (instanton) effect 
 of the $SU(2)$ gauge dynamics \cite{instanton}. 
The total effective superpotential is found to be 
\begin{eqnarray}
 W_{eff}= \lambda S \left[ Q_1 Q_2 \right] 
        + \frac{\Lambda^5}{\left[Q_1 Q_2 \right]} \; ,  
 \label{superpotential}
\end{eqnarray}
 where the second term is the dynamically generated superpotential
 and $ \Lambda $ is the dynamical scale of
 the $SU(2)_H$ gauge interaction. 
Note that the supersymmetric vacuum lies 
 at $ \langle S \rangle \rightarrow \infty$ and 
 $ \langle Q_1\rangle, \langle Q_2 \rangle \rightarrow 0$, 
 if only the F-term potential is considered. 
 
Next, let us consider the D-term potential. 
The gauged $U(1)_R$ symmetry is impossible
 in the globally supersymmetric theory, 
 since the generators of the $U(1)_R$ symmetry and supersymmetry
 do not commute with each other. 
On the other hand, in the supergravity theory the $U(1)_R$ symmetry 
 can be gauged as if it were a usual global symmetry 
 \cite{superconf,chamseddine}. 
However, it should be noticed that
 the Fayet-Iliopoulos term of the gauged $U(1)_R$ symmetry
 appears due to the symmetry of supergravity. 
This fact is easily understood
 by the standard formula for supergravity theories \cite{cremmer}. 
Using the generalized K\"ahler potential $G = K + \ln |W|^2$, 
 we have $D = \sum_i q_i (\partial G/ \partial z_i) z_i $, 
 where $q_i$ is the $U(1)_R$ charge of the field $z_i$. 
Note that the contribution from the superpotential leads to 
 the constant term,
 since the superpotential is holomorphic and has $U(1)_R$ charge 2. 
 
With the above particle contents, the D-term potential is found to be 
\begin{eqnarray}
 V_D = \frac{g_R^2}{2} \left( 
       4 S^\dagger S - Q_1^\dagger Q_1 - Q_2^\dagger Q_2 + 2 M_P 
       \right)^2  \; , 
\end{eqnarray}
 where $M_P = M_{pl}/\sqrt{8 \pi}$ is the reduced Planck mass, 
 $g_R$ is the $U(1)_R$ gauge coupling, 
 and the minimal K\"ahler potential, 
 $K= S^\dagger S + Q_1^\dagger Q_1 + Q_2^\dagger Q_2 $,  
 is assumed.
\footnote{
This assumption is justified by our result with $\Lambda \ll M_P$  
 which means that the $SU(2)_H$ gauge interaction
 is weak at the Planck scale. 
}
Note that the supersymmetric vacuum conditions 
required by the D-term potential 
and by the effective superpotential of Eq. (\ref{superpotential}) 
are incompatible. 
Therefore, supersymmetry is broken. 
This consequence remains correct, 
 if there is no other superfields which have negative $U(1)_R$ charges, 
 or if the other negatively charged superfields (if they exist) 
 have no vacuum expectation value.

Let us analyze the total potential in our model. 
Here, note that the cosmological constant should vanish. 
This requirement comes
 not only from the observations of the present universe 
 but also from the consistency of our discussion. 
Since it is not clear whether the superpotential discussed above 
 can be dynamically generated even in the curved space, 
 the space-time should be flat for our discussion to be correct. 
Note that we cannot take the usual strategy,
 namely, adding a constant term to the superpotential, 
 since such a term is forbidden by the $U(1)_R$ gauge symmetry. 
Therefore, it is a non-trivial problem whether we can obtain 
 the vanishing cosmological constant in our model. 

Assuming that the potential minimum lies on the D-flat direction 
 of the $SU(2)_H$ gauge interaction, 
 we take the vacuum expectation values such that 
 $ \langle S \rangle = s $ and 
 $\langle Q_i^\alpha  \rangle = v \delta_i^\alpha$, 
 where $i$ and $\alpha$ denote the flavor
 and $SU(2)_H$ indices, respectively. 
We can always make $s$ and $v$ real and positive 
 by symmetry transformations. 
The total potential is given by 
\begin{eqnarray}
   V(v,s) &=& e^K 
   \left[ \left( \lambda v^2 + s W \right)^2 
   + 2 v^2 \left( \lambda s - \frac{\Lambda^5}{v^4} + W \right)^2  
   -3 W^2  \right]    
 \label{potential}       \\ \nonumber  
   &+& \frac{g_R^2}{2} \left( 4 s^2 - 2 v^2 +2 \right)^2  \; , 
\end{eqnarray}
 where $K$ and $W$ are the K\"ahler potential and superpotential, 
 respectively, which are given by 
\begin{eqnarray}
\label{Kahler} 
 K &=& s^2 + 2 v^2  \; ,  \\
\label{superpot} 
 W &=& \lambda s v^2 + \frac{\Lambda^5}{v^2} \; . 
\end{eqnarray}
Here, all dimensionful parameters are taken to be dimensionless 
 with the normalization $M_P=1$. 
The first line in Eq. (\ref{potential}) comes from the F-term  
 (except for $W^2$ term) and the remainder is the D-term potential.  

Since the potential is very complicated, it is convenient 
 to make some assumptions for the values of parameters. 
First, assume that $g_R \gg \lambda, \Lambda$. 
Since the D-term potential is proportional to $g_R^2$
 and positive definite,
 the potential minimum is expected for $V_D$ to be small as possible. 
If we assume $s \ll 1 $ and $v \sim 1$, the potential can be rewritten as 
\begin{eqnarray}
 V \sim  e^2 \left( \lambda^2  - 3 \Lambda^{10}\right)  \; . 
\end{eqnarray}  
It is found that $\lambda \sim \sqrt{3} \Lambda^5$ is required 
 in order to obtain the vanishing cosmological constant. 
 
Let us consider the stationary conditions of the potential. 
Using the assumptions $s \ll 1$ and $v = 1+y $ ($|y| \ll 1$), 
 the stationary conditions can be expanded with respect to $s$ and $y$.  
Considering the relations $ g_R \gg \lambda \sim \Lambda^5$, 
we can expand the condition $\partial V/ \partial y =0$ and obtain 
\begin{eqnarray}
  y \sim s^2 - \frac{e^2 \lambda^2}{2 g_R^2}  \; . 
 \label{stationary1}
\end{eqnarray}
Using this result, 
we can also expand the expansion of the condition 
$\partial V/ \partial s  =0$ and obtain
\begin{eqnarray}
 s \sim \frac{ \lambda \Lambda^5} {8 \lambda^2 - \Lambda^{10}} \; .
 \label{stationary2}
\end{eqnarray}
By the numerical analysis, 
 the above rough estimation is found to be a good approximation. 
The result of numerical calculations is the following. 
\begin{eqnarray}
 y &\sim &  4.7 \times 10^{-3} \; ,
\label{y-value}   \\ 
 s &\sim &  6.8 \times 10^{-2} \; .  
\label{s-value}
\end{eqnarray}
Here, we used the values of $\Lambda =10^{-3}$, $ \lambda \sim 1.8 \;
\Lambda^5$ 
 and $g_R = 10^{-11}$. 
For these values of the parameters, 
 we can obtain the vanishing cosmological constant. 
Note that the numerical values of Eqs. (\ref{y-value}) 
and (\ref{s-value}) are almost independent of 
the actual value of $ \Lambda $, 
 if the condition $g_R \gg \Lambda^5$ is satisfied and 
 the ratio $\lambda/ \Lambda^5$ is fixed. 
This can be seen in the approximate formulae 
 of Eqs.(\ref{stationary1}) and (\ref{stationary2}). 
We can choose the value of $\Lambda $ 
in order to obtain a phenomenologically acceptable mass spectrum.

As a result of the above analysis,
 we can estimate the gravitino mass as
\begin{eqnarray}
  m_{3/2} = \langle e^{K/2} \; W  \rangle
  \sim  3.0 \times  \frac{\Lambda^5}{M_P^4}  \; .
\end{eqnarray}
This non-zero gravitino mass means that supersymmetry is really
 broken in the framework of the supergravity.
The gravitino mass contributes to the masses of scalar partners
 via the tree-level interactions of supergravity.
Note that there is another contribution,
 if scalar partners have non-zero $U(1)_R$ charges.
In this case, they also acquire the mass
 from the vacuum expectation value of the D-term,
 and it is estimated as
\begin{eqnarray}
  m_{D-term}^2 = g_R^2 \langle D \rangle q
  \sim \left( 7.3 \times \frac{\Lambda^5}{M_P^4} \right)^2 q \; ,
\end{eqnarray}
where $q$ is the $U(1)_R$ charge. 
This mass squared is always positive for the scalar fields
 with positive $U(1)_R$ charges. 
The mass is the same order of the magnitude of the gravitino mass. 
This is because $g_R$ is canceled out in the above estimation
 (see Eq. (\ref{stationary1})). 
Note that $m_{3/2}$ and $m^2_{D-term}$ is controled by the strong
 coupling scale of $SU(2)_H$ gauge theory $\Lambda$.

We give some comments. 
Our model has the same structure of the supersymmetry breaking model 
 with the anomalous $U(1)$ gauge symmetry \cite{anomalousU1}. 
In the model, the Fayet-Iliopoulos term is originated 
 from the anomaly of the  $U(1)$ gauge symmetry \cite{superstring}. 
On the other hand, in our model the origin of the term 
 is the symmetry of supergravity with the gauged $U(1)_R$ symmetry.  
The Fayet-Iliopoulos term appears even if
 the $U(1)_R$ gauge interaction is anomaly free.

The mediation of supersymmetry breaking to the visible
 sector is discussed in succeeding sections.
We address the highly non-trivial problem 
whether the anomaly cancellation of $U(1)_R$ can be done 
 when we include the visible sector superfields
 with positive semi-definite $U(1)_R$ charges \cite{chamseddine}. 
We discuss two explicit models. 
One is the model that the $U(1)_R$ anomalies are canceled out
 by the Green-Schwarz mechanism. 
We refer this model to the anomalous $U(1)_R$ model. 
The other is the model that the $U(1)_R$ symmetry is anomaly free.
We refer this model to the anomaly free $U(1)_R$ model.  

\section{Anomalous $U(1)_R$ model}

Before discussing the model in detail, 
it is instructive to review the Green-Schwarz mechanism
 in four dimensions \cite{green}. 
Suppose that we have a gauge symmetry $U(1)_X$ with
 mixed gauge anomalies of $U(1)_XG_i^2$,
 where $G_i$ denote other gauge groups.
The Lagrangian is not invariant under the $U(1)_X$ gauge transformation
  $A_{\mu} \to A_{\mu} + \partial_{\mu} \alpha(x)$:
\begin{equation}
\delta {\cal L}_{gauge} = \alpha(x) \frac{C_i}{8 \pi^2} {\rm tr} 
(F_i \tilde{F}_i),
\label{gaugepart}
\end{equation}
where $C_i$ are anomaly coefficients 
 and $F_i(\tilde{F_i})$ are the field strength tensors (its dual)
 of $G_i$. 
On the other hand, 
 there could be another anomaly at the Planck scale
 through the dilaton superfield $S$ with the $U(1)_X$
 transformation $S \to S + \frac{i}{2} \delta_{{\rm GS}} \alpha(x)$:
\begin{equation}
\delta {\cal L}_{{\rm GS}} = - \alpha(x)
 \frac{\delta_{{\rm GS}}}{2} k_i 
 \frac{1}{2} {\rm tr} (F_i \tilde{F_i}), \label{GS}
\end{equation}
where $\delta_{{\rm GS}}$ is the Green-Schwarz coefficient which 
 is calculated in the string theory \cite{superstring} as 
\begin{equation}
\delta_{{\rm GS}} = \frac{1}{192 \pi^2} {\rm tr} q_i.
\end{equation}
Here, $q_i$ are R-charges of the fermionic components of the superfields. 
$k_i$ are the Kac-Moody levels of $G_i$. 
Note that the Kac-Moody levels are integers for non-Abelian gauge
 groups but are not necesarry integers for Abelian groups. 
Therefore, we can assume the value of the Kac-Moody levels
 for Abelian groups appropiately. 
All the Kac-Moody levels must have the same sign,
 since all the gauge couplings are generated by
\begin{equation}
\frac{1}{g_i^2} = k_i \langle S \rangle,
\end{equation}
where $\langle S \rangle$ is the vacuum expextation value
 of the dilaton superfield. From Eqs. (\ref{gaugepart})
 and (\ref{GS}), 
 the anomaly cancellation conditions lead to
 the Green-Schwarz relations:
\begin{equation}
\frac{C_i}{k_i} = 2 \pi^2 \delta_{{\rm GS}}.
\end{equation}
This relation should be satisfied for any $i$. 
It follows from this relation that all of $C_i$ has the same sign,
 since all of $k_i$ has the same sign.

Now, we discuss the model in detail. 
We consider the MSSM with all the Yukawa couplings
 but the $\mu$-term. 
The relevant anomaly coefficients are listed below.
\begin{eqnarray}
C_H &=& \frac{1}{2} (q_1 + q_2) + 2, \label{hidden} \\
C_3 &=& \frac{3}{2} (2q + u + d) + 3, \label{C3} \\
C_2 &=& \frac{3}{2} (3q + l) + \frac{1}{2} (h + \bar{h}) + 2, 
\label{C2} \\
C_Y &=& 3 (\frac{1}{6} q + \frac{4}{3} u + \frac{1}{3} d + 
\frac{1}{2} l + e) + \frac{1}{2} (h + \bar{h}), \label{C1} \\
C_{YY} &=& 3 (q^2 -2 u^2 + d^2 - l^2 + e^2) + h^2 - \bar{h}^2 = 0, 
\label{C11} \\
C_g &=& 2 (q_1 + q_2) + s + 3 (6q + 3u + 3d + 2l + n + e) + 
2 (h + \bar{h}) - 6, \label{Cg} \\ 
C_R &=& 2 (q_1^3 + q_2^3) + s^3 + 3 (6q^3 + 3u^3 + 3d^3 + 
2l^3 + n^3 + e^3) 
+ 2 (h^3 + \bar{h}^3) + 18, \label{CR}
\end{eqnarray}
where $q_1, q_2$ and $s$ are R-charges of the fermionic components 
 for $Q_1, Q_2$ and $S$ in the hidden sector, and 
 $q, u, d, l, n, e, h$, and $\bar{h}$ are those for the chiral superfields
 $Q({\bf 3},{\bf 2},\frac{1}{6})$, 
 $\bar{U}(\bar{{\bf 3}}, {\bf 1}, -\frac{2}{3})$, 
 $\bar{D}(\bar{{\bf 3}}, {\bf 1}, \frac{1}{3})$, 
 $L({\bf 1}, {\bf 2}, -\frac{1}{2})$, 
 $N({\bf 1}, {\bf 1}, 0)$
 $E({\bf 1}, {\bf 1}, 1)$, 
 $H({\bf 1}, {\bf 2}, \frac{1}{2})$, 
and 
 $\bar{H}({\bf 1}, {\bf 2}, -\frac{1}{2})$, respectively. 
The representations and charges in the parenthesis are those 
 under $SU(3)_C \times SU(2)_L \times U(1)_Y$. 
The coefficients $C_H, C_3, C_2, C_Y, C_{YY}, C_g$ and $C_R$ represent 
 the anomaly coefficients for 
 $U(1)_R(SU(2)_H)^2$, $U(1)_R(SU(3)_C)^2$, $U(1)_R(SU(2)_L)^2$, 
 $U(1)_R(U(1)_Y)^2$, $(U(1)_R)^2U(1)_Y$, $U(1)_R$ and $(U(1)_R)^3$, 
 respectively.
Here we simply assume a family independent charge assignment. 
Note that the gravitino contribution for the mixed gravitational
 anomaly is $-21$ times that of a gaugino,
 while the gravitino contribution for the $U(1)_R$
 gauge anomaly is three times that of a gaugino \cite{gravitino},
 and the dilatino contributes ($-1$) to both $C_g$ and $C_R$.  
Note also that the anomaly coefficient $C_{YY}$ has to
 vanish identically, since it cannot be cancelled by
 the Green-Schwarz mechanism.

As mentioned earlier, 
all the Yukawa coupling are included. 
\begin{equation}
W = y_u Q \bar{U} H + y_d Q \bar{D} \bar{H} 
+ y_e L E \bar{H} + y_n L N H. 
\end{equation}
The resulting conditions for R-charges are
\begin{eqnarray}
q + u + h &=& -1, \label{up} \\
q + d + \bar{h} &=& -1, \label{down} \\
l + e + \bar{h} &=& -1, \label{electron} \\
l + n + h &=& -1. \label{neutrino}
\end{eqnarray}
Furthermore, we need the Yukawa coupling in the hidden sector
\begin{equation}
\label{hiddensp}
W = \lambda S[Q_1Q_2],
\end{equation}
which leads to the condition 
\begin{equation}
\label{spcon}
s + q_1 + q_2 = -1. 
\end{equation}
One can immediately see that $C_3 = C_2 = C_Y = 0$ and 
 Eqs. (\ref{up}), (\ref{down}) and (\ref{electron}) are incompatible, 
 since $C_Y + C_2 - 2C_3 = - 6$. 
Therefore we have to use the Green-Schwarz mechanism to cancel 
 the anomalies.

Taking into account the simple assumption
 $k_2 = k_3$, we have only two solutions.
One solution consists of all integer charges:
\begin{eqnarray}
&& Q_1 = Q_2 = -2, \quad S = 6, \\
&& Q = \bar{U} = \bar{D} = L = N =E = 0, 
\quad H = \bar{H} = 2,
\end{eqnarray}
where these are the charges of superfields. 
The corresponding anomaly coefficients are
\begin{eqnarray}
&& C_3 = C_2 = -3, \quad C_Y = -9, \quad C_{YY} = 0, \\
&& C_H = -1, \quad C_R = -9, \quad C_g = -57.
\end{eqnarray}
Note that all the non-trivial anomaly coefficients have 
 negative signs. 
Therefore, we assume that all the Kac-Moody levels are positive
 with negative vacuum expectation value of the dilaton superfield. 
We can freely choose the positive values of $k_R$ and $k_g$
 to satisfy the Green-Schwarz relation \cite{green}.
On the other Kac-Moody levels the following discussions
 are required.

The Green-Schwarz relation
\begin{equation}
\frac{C_Y}{k_Y} = \frac{C_2}{k_2} = \frac{C_3}{k_3} = 
\frac{C_H}{k_H} = 2 \pi^2 \delta_{GS} = 
\frac{2 \pi^2 C_g}{192 \pi^2} = \frac{C_g}{96}
\end{equation}
is satisfied, if we introduce 39 gauge singlets of
 vanishing R-charge with $k_Y  = 9, k_2 = k_3 = 3$ and $k_H = 1$. 
These values of Kac-Moody levels tell us 
 the gauge coupling relations at the Planck scale, namely, 
\begin{equation}
\alpha_3 = \alpha_2 = 3 \alpha_Y, \quad \alpha_H = 3 \alpha_3.
\end{equation}
This coupling unification can be easily accomplished 
 by introducing extra massive particles
 which change the running of the gauge
 coupling constants and
 does not affect the anomaly cancellation conditions. 
In fact, the gauge coupling unification is realized,
 if we introduce extra massive particles so that
 the $SU(3)_C$ gauge coupling almost does not run and
 the $SU(2)_L$ gauge coupling is asymptotically non-free.

Let us perform the potential analysis.
We assume the $SU(2)_H$ D-flat condition. 
The scalar potential in the present case is the same one 
 in our previous letter except for 
 the D-term part\footnote{Although the Fayet-Iliopoulos term
 due to the anomaly of $U(1)_R$ should be 
 included, but its magnitude is suppressed
 compared to that of the gauged $U(1)_R$ symmetry.
 Hence, we simply neglect it.}
\begin{eqnarray}
V(v,s) &=& 
 e^K \left[ \left( \lambda v^2 + s W \right)^2 + 
 2 v^2 \left( \lambda s - \frac{\Lambda^5}{v^4} + W \right)^2 
 - 3W^2 \right] \nonumber \\
 && + \frac{g_R^2}{2} \left( 6s^2 - 4v^2 + 2 \right)^2,
\end{eqnarray}
where $K$ and $W$ are the K\"ahler potential and the superpotential,
 respectively, which are given by Eqs. (\ref{Kahler}) 
 and (\ref{superpot}).
We assume $\lambda, \Lambda \ll g_R, v = \frac{1}{\sqrt{2}} + y$
 with $y \ll 1$ and $s \ll 1$, and
 expand the potential with respect to $y$ and $s$. 
The approximate formulae of the stationary conditions
 $\partial V/\partial y = 0$
 and $\partial V/\partial s = 0$ are
\begin{eqnarray}
y &\sim& \frac{3}{2\sqrt{2}} s^2 - 
 \frac{e( 3\lambda^2 - 16 \Lambda^{10})}{32 \sqrt{2} g_R^2}, \\
s &\sim& \frac{10 \lambda \Lambda^5}{9 \lambda^2 -32 \Lambda^{10}}.
\end{eqnarray}
These formulae are correct within the order of magnitude,
 since the convergence of the expansion is not so good. 
The numerical calculation gives
\begin{eqnarray}
v &\sim& \frac{1}{\sqrt{2}} + 0.019, \\
s &\sim& 0.13,
\end{eqnarray}
where we used the parameters,
 $\Lambda = 10^{-3}$, $\lambda = 7.3 \Lambda^5$ and
 $g_R^2 = 10^{-11}$. 
We have checked that
 the cosmological constant can be fine-tuned to zero
 by adjusting $\lambda$ for fixed $\Lambda$.

Next, we estimate the spectrum of supersymmetry breaking masses. 
The gravitino mass is given by
\begin{equation}
m_{3/2} = \langle e^{K/2} W \rangle \sim 4.1 
 \times \frac{\Lambda^5}{M_P^4}.
\end{equation}
This gravitino mass contributes to the masses of scalar partners
 via the tree-level interaction of supergravity. 
Since R-charges for $Q, \bar{U}, \bar{D}, L, N, E$ vanish,
%
\begin{equation}
m_i^2 \sim m_{3/2}^2 \quad (i = Q, \bar{U}, \bar{D}, L, N, E).
\end{equation}
On the other hand, since $H$ and $\bar{H}$ have R-charge 2, 
%
\begin{eqnarray}
m^2_{H,\bar{H}} &\sim& m_{3/2}^2 + 2g_R^2 \langle D \rangle, \nonumber \\
 &\sim& m_{3/2}^2 + 2 \left( 8.2 \frac{\Lambda^5}{M_P^4} \right)^2
 \sim {\cal O}(m_{3/2}^2).
\end{eqnarray}
The contribution from $\langle D \rangle \ne 0$ is 
the same order of the gravitino mass. 
%
%

The gaugino mass can be generated in two different ways. 
In the case that the gauge kinetic function is trivial,
 the anomaly mediation contribution dominates as
\begin{equation}
m_{\lambda_i} = \frac{\beta(g_i^2)}{2g_i^2}m_{3/2},
\end{equation}
where $\beta(g_i^2)$ is the beta function for the standard model
 gauge groups.\footnote{More precisely,
 there is a correction of the same order of magnitude,
 see Ref. \cite{BMP}.}
In the case that the gauge kinetic function is non-trivial
 and includes the term $S^2 [Q_1Q_2]^3/M_P^8$, for example,
 the gaugino masses are 
\begin{equation}
m_{\lambda_i} \sim \frac{1}{2^3} m_{3/2}.
\end{equation}
In both cases, there could be further contribution from
 the vacuum expectation value of F-component of the dilaton \cite{ADM}. 
The experimental bound on gaugino masses in the MSSM \cite{PDG}
 determines the order of the gravitino mass as
 10 TeV in both cases.

Here, we comment on the $\mu$-term. 
We can have the higher dimensional interaction
\begin{equation}
\label{muterm}
W = \kappa \frac{S[Q_1Q_2]^2}{M_P^4} H\bar{H}
\end{equation}
which respects all symmetries, 
where $\kappa$ is a dimensionless constant.
Plugging the vacuum expectation value of $S$, $Q_1$ and $Q_2$
 into Eq. (\ref{muterm}),
 we obtain $\mu \sim \kappa \langle S \rangle$.
By adjusting $\kappa \ll 1$, the electroweak scale is generated.

\section{Anomaly free $U(1)_R$ model}
In the last section,
 we discussed the scenario that supersymmetry breaking effect
 is mediated to the MSSM as the visible sector. 
The Green-Schwarz mechanism is used to cancel gauge anomalies
 of $U(1)_R$ and the simple solution was found. 
However the Green-Schwarz mechanism requires the dilaton
 field, and
this leads to new difficult problems such as the dilaton
 stabilization and so on \cite{ADM,dilaton}. 
Therefore, it is desirable to consider the case that
 $U(1)_R$ is anomaly free.
As mentioned earlier,
 it is impossible to cancel all the gauge anomaly for $U(1)_R$
 in the MSSM. 
In this section, 
 we consider the supersymmetric SU(5) GUT instead of the MSSM
 as the visible sector. 
We have found that
 all the gauge anomalies can be cancelled by introducing
 $SU(5) \times SU(2)_H$ gauge singlets with non-trivial R-charges.

Suppose that we have $N$ of $SU(5) \times SU(2)_H$ gauge singlets
 with R-charge 2. 
The anomaly cancellation conditions are
\begin{eqnarray}
&&U(1)_R \left[SU(5) \right]^2 : 3 \left(\frac{1}{2} \bar{f}
 + \frac{3}{2} a \right) + \frac{1}{2} (h + \bar{h})
 + 5 \sigma + 5 = 0, \\
&&\left[U(1)_R \right] : 2(q_1 + q_2) + s + 3(5 \bar{f} + 10a + n)
 + 5 (h + \bar{h}) + 24 \sigma + N + 7 = 0, \\
&&\left[U(1)_R \right]^3 : 2(q_1^3 + q_2^3) + s^3
 + 3(5 \bar{f}^3 + 10 a^3 + n^3)
 + 5(h^3 + \bar{h}^3) + 24 \sigma^3 + N + 31 = 0,
\end{eqnarray}
where $\bar{f}, a, n, \sigma, h$ and $\bar{h}$ are
 the R-charges of the fermionic component for the superfields
 $\bar{F}({\bf \bar{5}})$, $A({\bf 10})$, $N({\bf 1})$,
 $\Sigma({\bf 24})$, $H({\bf 5})$ and
 $\bar{H}({\bf \bar{5}})$, respectively. 
The representations in the parethesis are those of SU(5).
The superpotential of Eq. (\ref{hiddensp}),
 and Yukawa couplings
\begin{equation}
W \sim AAH + A\bar{F}\bar{H} + N\bar{F}H
\end{equation}
give the conditions of Eq. (\ref{spcon}) and
\begin{eqnarray}
2a + h &=& -1, \\
\bar{f} + a + \bar{h} &=& -1 \\
n + \bar{f} + h &=& -1.
\end{eqnarray}
We have a simple solution, in the case of $N = 36$ as
\begin{eqnarray}
&& \bar{F} = A = N = 0, \quad H = \bar{H} = 2, \quad
 \Sigma = 1, \nonumber \\
&& Q_1 = 0, \quad Q_2 = -2, \quad S = 4,
\end{eqnarray}
where these are the charges of superfields. 
Note that $Q_1$ and $Q_2$ have different charges now. 
This modification makes possible to have the special
 charge assignment with all positive charges except for
 the fields which couple to $SU(2)_H$ gauge bosons.

The potential analysis is the same in Ref. \cite{kitazawa}
 as long as we assume that the potential is
 along the $SU(2)_H$ D-flat direction. 
The gravitino mass is estimated as
\begin{eqnarray}
  m_{3/2} = \langle e^{K/2} \; W  \rangle
  \sim  \frac{\Lambda^5}{M_P^4}  \; .
\end{eqnarray}
The gravitino mass contributes to the masses of scalar partners
 via the tree-level interactions of supergravity.
Note that there is another contribution,
 if scalar partners have non-zero $U(1)_R$ charges.
In this case, they also acquire the mass
 from the vacuum expectation value of the D-term,
 and it is estimated as
\begin{eqnarray}
  m_{D-term}^2 = q g_R^2 \langle D \rangle
  \sim \left(\frac{\Lambda^5}{M_P^4} \right)^2 q \; ,
\end{eqnarray}
where $q$ is the $U(1)_R$ charge.
This mass squared is always positive
 for the scalar partners with positive $U(1)_R$ charges.
The mass is the same order of the magnitude of the gravitino mass.

The gaugino mass can be generated in two different ways. 
In the case that the gauge kinetic function is trivial,
 the anomaly mediation contribution dominates as
\begin{equation}
m_{\lambda_i} = \frac{\beta(g_i^2)}{2g_i^2} m_{3/2}.
\end{equation}
On the other hand, in the case that the gauge kinetic
 function is non-trivial and includes the higher dimensional term
 $S([Q_1Q_2])^2/M_P^5$
\footnote{
 The higher dimensional term $ S ([Q_1 Q_2])^2 / M_P^5 $
 in the gauge kinetic function
 can be forbidden to all orders by the discrete symmetry.
 }, for example, the gaugino masses are 
\begin{equation}
m_{\lambda_i} \sim m_{3/2}.
\end{equation}
Considering the experimental bound on gaugino masses
 in the MSSM \cite{PDG},
 the gravitino mass is taken to be of the order of 10 TeV or 1 TeV
 in the former case or the latter case, respectively.
This phenomenological constraint requires
 the dynamical scale of the $SU(2)_H$ gauge interaction
 to be of the order of $10^{15}$ GeV for both cases.
This also means that $\lambda$ is extremely small,
 $\lambda \sim 10^{-15}$.
Note that the requirement of this fine-tuned value of $\lambda$
 is the result from the fine-tuning for vanishing cosmological constant.
Furthermore,
 this small Yukawa coupling is consistent with the above discussion
 in the following sense.
Since $S$ has the vacuum expectation value, the mass for $Q_i$
 is generated through the Yukawa coupling in Eq.(\ref{superpotential}).
The relation $\lambda \langle S \rangle \ll \Lambda $ is needed
 not to change our result from the $SU(2)_H$ gauge dynamics.

Here, we comment on the Higgs potential.
We would like to discuss whether it is possible
 to construct the higher dimensional operators
 that realize the gauge symmetry breaking as
 $SU(5) \to SU(3)_C \times SU(2)_L \times U(1)_Y \to
 SU(3)_C \times U(1)_{em}$ and
 also realize the doublet-triplet splitting. 
We can explicitly write down the following interactions
 respecting all symmetries.\footnote{More precisely,
 there are other terms of contracting the gauge indices
 in different ways, but we simply write the representative.}
\begin{eqnarray}
W &\sim& \frac{\lambda_1}{M_P} \left[ Q_1Q_2 \right] \bar{H}H
 + \frac{\lambda_2}{M_P^5} \left[ Q_1Q_2 \right]^2
 \bar{H} \Sigma^2 H  \nonumber \\
 && + \frac{\lambda_3}{2M_P^4} S \left[ Q_1Q_2 \right]^2
 {\rm tr}(\Sigma^2) 
 + \frac{\lambda_4}{4M_P^3} \left[ Q_1Q_2 \right] 
 {\rm tr}(\Sigma^4),
\end{eqnarray}
where $\lambda_i (i = 1 \sim 4)$ are constants. 
By choosing $\lambda_i$ appropriately,
 the first two terms can realize the doublet-triplet
 splitting, while the last two terms can realize
 the gauge symmetry breaking.

Unfortunately, there exists unwanted operators, for examples,
\begin{eqnarray}
\frac{1}{M_P^3} \left[ Q_1Q_2 \right] (H\bar{F})^2
 &\to& \frac{1}{M_P} (H\bar{F})^2, \nonumber \\
\frac{1}{M_P^5} \left[ Q_1Q_2 \right]^3 S^2 &\to& M_P S^2, \\
\frac{1}{M_P^{10}} \left[ Q_1Q_2 \right]^5 S^3 &\to& S^3. \nonumber
\end{eqnarray}
Although the first term seems to be harmless,
 the other terms are dangerous since supersymmetry
 is restored if these operators are present. 
However, this problem is not specific to our model,
 but inevitable and generic in supergravity models.

\section{Summary and Discussion}
In this paper, we have shown that it is possible to construct
 simple and phenomenologically viable models of dynamical
 supersymmetry breaking with gauged $U(1)_R$ symmetry. 
Supersymmetry breaking occurs by the interplay
 between the dynamically generated superpotential
 and the Fayet-Iliopoulos term which appears
 due to the symmetry of supergravity. 
The cosmological constant can be fine-tuned to vanish
 at the minimum of the potential. 
What is the most non-trivial in this class of models is
 the anomaly cancellations for $U(1)_R$. 
We have presented two explicit models with anomaly cancellations
 as the visible sector.

One is the MSSM with all the Yukawa couplings and
 without the $\mu$-term. 
Anomalies are cancelled by the Green-Schwarz mechanism. 
We have found a quite simple solution of R-charge assignments for
 matter superfields in our model compared to the models
 in Ref. \cite{chamseddine}. 
Our solution consists of all integer charges. 
We have discussed the gauge coupling unification which follows from
 the Green-Schwarz relations. 
The gauge coupling unification is easily accomplished
 by introducing extra massive particles. 
The spectrum of supersymmetry breaking masses has been estimated
 and turned out to be phenomenologically viable as follows. 
The gravitino mass is of order 10 TeV. 
The scalar masses are the same order of the gravitino
 mass. 
The gaugino masses are a few orders
 smaller than the gravitino mass. 
We have also shown that it is possible to have
 the $\mu$-term. 
Unfortunately, we have to introduce the dilaton superfield to
 cancel anomalies in this model. 
This leads to new difficult problems such as
 the dilaton stabilization and so on \cite{ADM,dilaton}.

The other is the supersymmetric $SU(5)$ GUT. 
In this model, $U(1)_R$ anomalies are cancelled by introducing
 the $SU(5)$ gauge singlet superfields.
We have also found a quite simple solution of R-charge assignment. 
Unlike the anomalous $U(1)_R$ model in this paper and
 the anomalous $U(1)$ model in Ref. \cite{anomalousU1},
 the dilaton is not necessary since the gauged $U(1)_R$ is
 anomaly free. 
Therefore, there is no dilaton stabilization problem.
We have estimated the mass spectrum of supersymmetry breaking
 and found that these models are phenomenologically viable. 
The gravitino mass is of the order 1 TeV or 10 TeV
 depending on the form of the gauge kinetic function. 
The scalar masses are the same order of the gravitino mass. 
The gaugino masses are a few orders smaller
 than the gravitino mass in the case that
 the gauge kinetic function is trivial,
 while the same order of it in the case that
 the gauge kinetic function have higher dimensional terms.
We have shown that it is possible to have the Higgs potential
 for the doublet-triplet splitting and
 the GUT gauge symmetry breaking.

It is interesting to apply our mechanism to the phenomenology
 of the anomaly mediation scenarios. 
In this scenarios, if the visible sector consists of the MSSM,
 the slepton mass squared becomes negative
 because the beta function coefficients of $SU(2)_L \times U(1)_Y$
 are negative. 
This problem is easily avoided
\footnote{
For related works, see \cite{slepton}}
 by using our mechanism
 since the contribution to the mass squared from
 the vacuum expectation value of $U(1)_R$ D-term
 is always positive for the positively R-charged field. 
Unfortunately, this situation is not realized in our models
 since R-charges of the matter fields are zero. 
However, it would be possible to construct the models
 in which lepton superfields have positive charges.

Finally, note that our model is effectively reduced
 to the well known Polonyi model which is the simplest
 supersymmetry breaking model at the tree-level. 
One can understand this fact by freezing the $Q_i$ with its
 vacuum expectation values in the superpotential of
 Eq. (\ref{superpotential}).
Here, we would like to emphasize that
 the Polonyi model is derived as the effective theory of our model. 
The Polonyi model does not specify the dynamics of
 supersymmetry breaking. 
Also, unlike the Polonyi model,
 there is no dimensionful parameters
 other than the Planck scale in our model. 
They are induced dynamically from the Planck scale.

\acknowledgments
This work was supported in part 
 by the Grant-in-aid for Science and 
 Culture Research form the Ministry of Education, 
 Science and Culture of Japan (\#11740156, \#3400, \#2997). 
N.M. and N.O. are supported by the Japan Society for 
 the Promotion of Science for Young Scientists.
%

%
\end{document}